\documentclass[twocolumn,prl,nobalancelastpage,aps,10pt]{revtex4-1}
\usepackage{graphicx,bm,times}
\usepackage{textcomp}
\usepackage{natbib} 
\usepackage{etoolbox}

\usepackage{enumitem}

\begin{document}

\title{Exploring the Frontiers: Challenges and Theories Beyond the Standard Model}

\author{Dhananjay Saikumar}

\affiliation{University of St Andrews}
\renewcommand{\thesubsection}{(\alph{subsection})}

\begin{abstract}
Quantum Field Theory (QFT) forms the bedrock of the Standard Model (SM) of particle physics, a powerful framework that delineates the fundamental constituents and interactions of the universe. However, the SM's narrative is incomplete, as it conspicuously fails to account for several empirical phenomena that challenge our current understanding of particle physics. This review meticulously examines three paramount anomalies that elude SM predictions: the elusive nature of dark matter, the Higgs boson's anomalously low mass, and the intricate puzzle of neutrino masses. Through a critical analysis, it delves into the forefront of theoretical advancements proposed to bridge these gaps, notably the Weakly Interacting Massive Particles (WIMPs), Supersymmetry (SUSY), and the intriguing hypothesis of right-handed neutrinos. By synthesizing current research and theoretical models, this review not only elucidates these profound mysteries but also underscores the imperative for a more comprehensive and unified theory of particle physics, setting the stage for future discoveries and theoretical breakthroughs.
\end{abstract}
\maketitle

\section{1. INTRODUCTION}
The Standard Model (SM) of particle physics represents the pinnacle of our understanding of the universe's fundamental forces, encapsulating the weak, strong, and electromagnetic forces, albeit excluding gravity. It systematically classifies elementary particles by their charges, thereby delineating their interactions with these fundamental forces. Charges, intrinsic properties of elementary particles, dictate their susceptibility to specific forces: color, electric, and weak charges are pertinent to strong, electromagnetic, and weak forces, respectively [1]. The SM's framework was established in the mid-1970s, catalyzed by the discovery of quarks [2]. Its validity has been corroborated by subsequent discoveries, including the top quark (1995), tau neutrino (2000), and the Higgs boson (2012), each reinforcing the SM's credibility [3].

\begin{figure}[h]
\centering
\includegraphics*[width=0.8\linewidth,clip]{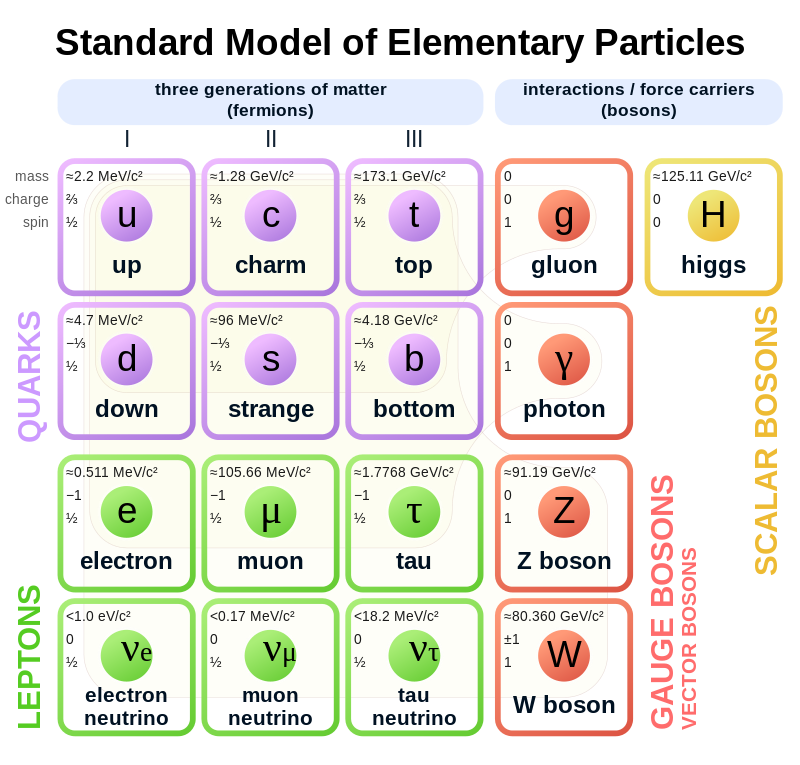}
\caption{Standard Model of Elementary Particles.}
\end{figure}

Within the SM, there are 17 elementary particles, divided into 12 fermions and 5 bosons. Fermions, with a spin of \(\frac{1}{2}\), constitute matter's building blocks and are further subdivided into quarks and leptons. Quarks, which are both electrically and color charged, partake in electromagnetic and strong interactions. Conversely, leptons, devoid of color charge, comprise electron varieties and neutrinos; electrons engage in electromagnetic and weak interactions, while neutrinos interact exclusively via the weak force. The interactions among these particles are mediated by bosons (spin-1 particles): the photon (\(\gamma\)) for electromagnetic interactions, gluons (\(g\)) for strong interactions, and the weak bosons (\(W^-\), \(W^+\), \(Z^0\)) for the weak force [4].

The Large Hadron Collider (LHC) was conceived to investigate the Electroweak Symmetry Breaking (EWSB), a phenomenon explaining the mass disparity between the massless photon and the more massive weak bosons [5]. The Higgs boson, a particle emanating from the scalar Higgs field that permeates the universe, was predicted in 1964. It interacts with most fundamental particles, imparting mass through the Higgs mechanism, as confirmed by the ATLAS and CMS experiments in 2012 [6-9].

\begin{figure}[h]
\centering
\includegraphics*[width=0.8\linewidth,clip]{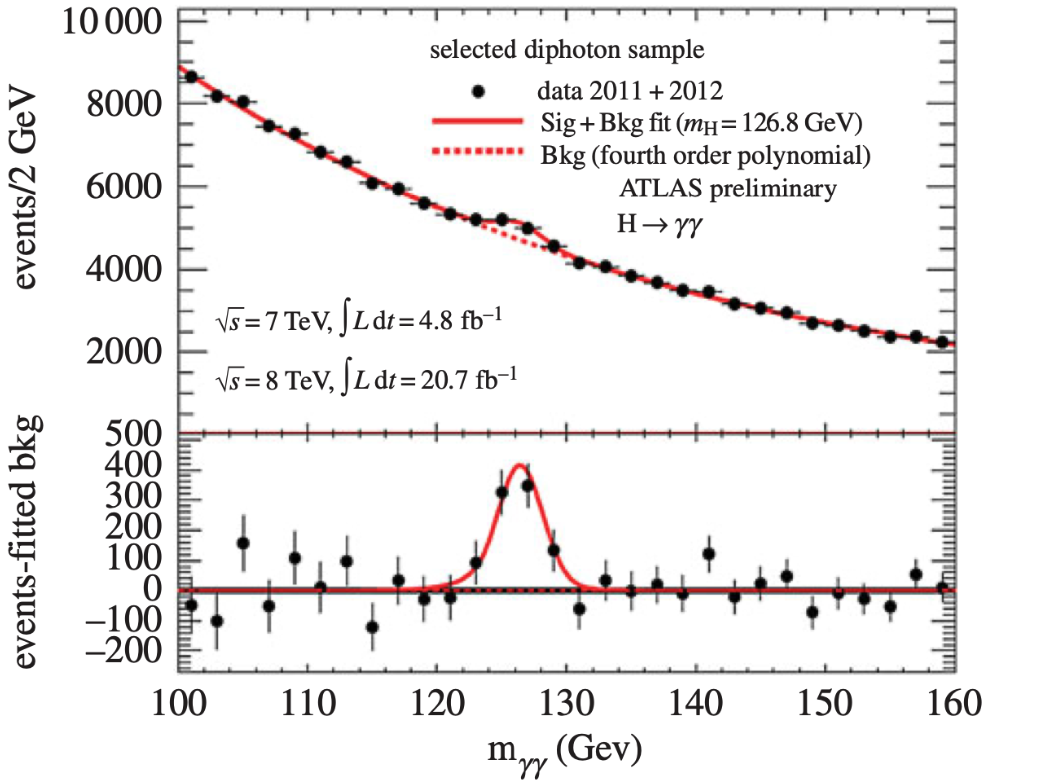}
\caption{Invariant mass distribution of the Higgs di-photon (particle decaying into two photons) candidates at ATLAS. The top inset dis- plays the mass distribution of the selected candidates. The bottom inset represents the signal fitted relative to the background. Ref [10]}
\end{figure}

Despite the SM's accurate predictions across numerous processes, cosmological observations, particularly from the Planck satellite, suggest a universe composition of 68\% dark energy and 32\% matter, with the SM accounting for merely about 5\% of this composition [10]. This disparity underscores the need for physics beyond the Standard Model (BSM), aiming to elucidate phenomena inadequately explained by the SM, including:
\begin{enumerate}[itemsep=0pt, parsep=0pt]
\item Dark Matter (DM)
\item The unanticipated lightness of the Higgs boson
\item Neutrino mass mechanisms
\item Gravitational interactions
\item The matter-antimatter asymmetry
\item The nature of dark energy
\end{enumerate}

This review endeavors to overview these BSM phenomena and their respective theoretical propositions, thereby expanding our comprehension of the universe beyond the current paradigms set by the Standard Model.

\section{2. DARK MATTER}
As previously discussed, the universe's matter density comprises approximately 32\% of its total composition, with 5\% being baryonic (ordinary) matter and 27\% attributed to dark matter (DM). The existence of dark matter was initially inferred by Jan Oort through the anomalous velocity distribution of stars relative to their distance from the Milky Way's center. Oort hypothesized the necessity of significant unseen mass to account for the stars' orbital velocities, despite acknowledging potential errors in velocity measurements [12]. Subsequently, in a pivotal study four decades later, the rotation curves of 60 spiral galaxies were analyzed through Doppler shift measurements of hydrogen and helium gas clouds, challenging the expected Newtonian/Keplerian dynamics [13]. This analysis revealed flat rotation curves, which implies that the enclosed mass \(m(r)\) is proportional to the radius \(r\), \(m(r) \propto r\), suggesting the presence of a significant non-luminous mass component and thereby bolstering the concept of dark matter.

Dark matter is postulated to consist of undiscovered particles that interact with baryonic matter primarily through gravity and potentially the weak force. A leading candidate for this particle is the Weakly Interacting Massive Particle (WIMP) [17]. Efforts to detect dark matter particles bifurcate into direct and indirect searches, with the latter discussed subsequently. The Large Underground Xenon (LUX) experiment, employing 370 kg of liquid xenon, exemplifies a direct detection strategy, aiming to observe WIMP-induced nuclear recoil through scintillation or ionization signals [18]. Initial findings in 2014 did not identify significant nuclear recoil events beyond background levels, thereby constraining low-mass WIMP scenarios [19]. The ongoing development and future operation of an advanced LUX experiment, featuring a 7-ton liquid xenon detector set to begin in 2022, generates considerable anticipation for the potential discovery of heavy WIMP candidates. Should this endeavor not yield the anticipated results, the scientific community may need to explore alternative models for dark matter.

\begin{figure}[h]
\centering
\includegraphics*[width=0.8\linewidth,clip]{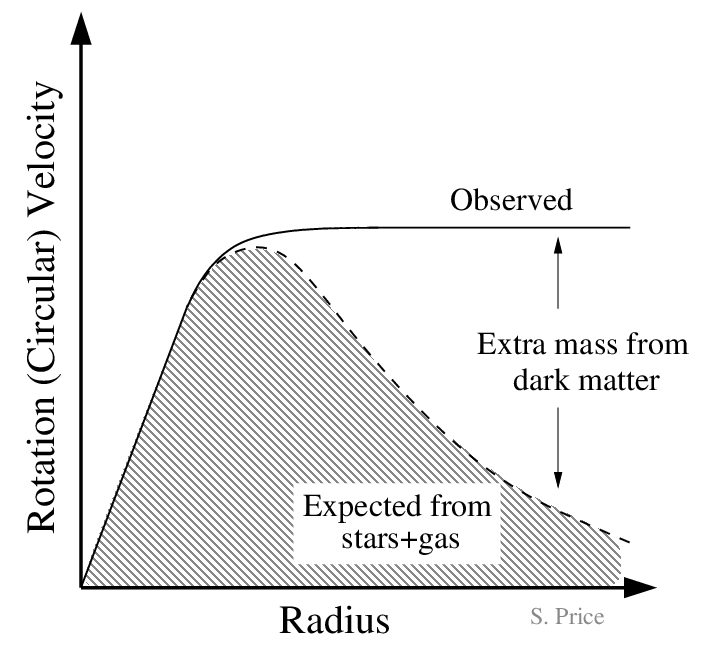}
\caption{Rotation velocities of HI regions.}
\end{figure}

\section{3. THE LIGHTNESS OF THE HIGGS BOSON}

Radiative correction is a critical method for extending the Standard Model's (SM) predictions to higher energy scales [1]. This approach is necessary because at elevated energy levels, multiple transition modes from an initial to a final state can occur, involving quantum loops and self-interactions that must be comprehensively accounted for. The Higgs boson mass is subject to adjustment as follows:
\begin{equation}
m_H^2 \approx m_{\text{bare}}^2 + \Lambda^2
\end{equation}

where $m_H$ denotes the observed mass of the Higgs boson (125 GeV) [20], $m_{\text{bare}}$ represents the intrinsic mass of the Higgs boson absent any interactions, and $\Lambda$ signifies the SM's validity cutoff energy, potentially reaching the Grand Unification Theory scale (around $10^{16}$ GeV), where all forces except gravity unify [21]. To reconcile the observed $m_H$ with theoretical models, a nearly exact cancellation between $m_{\text{bare}}$ and $\Lambda$ is required, leading to what is known as the fine-tuning problem, a scenario viewed with skepticism due to its perceived improbability [22].

\begin{figure}[h]
\centering
\includegraphics*[width=1\linewidth,clip]{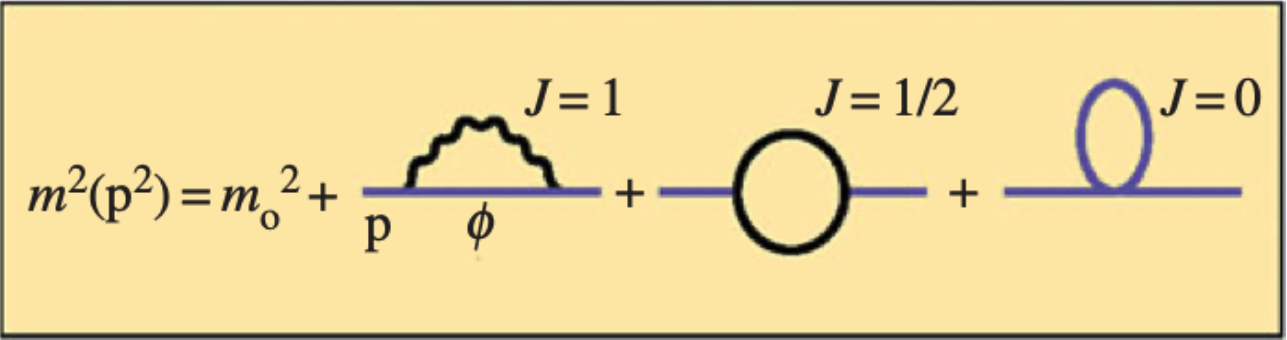}
\caption{Radiative corrections to the mass of the Higgs boson from multiple quantum loops (self-interactions) containing virtual fermions or bosons which become more accessible at higher energy scales.}
\end{figure}

The significance of radiative corrections, evidenced by their successful predictions in various contexts such as the mass of the top quark and the muon's lifetime, should not be underestimated [23]. If $m_{\text{bare}}$ aligns closely with $m_H$, it suggests that $\Lambda$ should be within the 1-10 TeV range, indicating the emergence of new physics at this scale. This raises a critical question: what mechanism ensures the Higgs boson's mass remains protected?

Two theories, Supersymmetry (SUSY) and the Composite Higgs model, propose answers to this dilemma. SUSY, in particular, introduces a novel symmetry, suggesting each fermion or boson has a corresponding super-partner with identical attributes except for spin [24]. For instance, squarks (spin 0) and sleptons (spin 0) are super-partners to quarks and leptons, respectively, while the Higgsino (spin \( \frac{1}{2} \)) corresponds to the Higgs boson. SUSY theory posits that radiative corrections to the Higgs boson's mass are counterbalanced by those of the Higgsino, effectively isolating the $m_{\text{bare}}$ term [24]. The observed mass of the Higgs boson being below 135 GeV has been touted as a prediction of SUSY, influencing the LHC's design and objectives [25].

SUSY also introduces interactions that potentially violate baryon and lepton number conservation, mitigated by a new quantum number, R-parity, distinguishing super-partners (R-parity = 1) from SM particles (R-parity = 0) [26]. This framework implies that the lightest super-partner (LSP) is stable and electrically neutral, rendering it a viable dark matter candidate. The search for the LSP at the LHC involves detecting the decay products of super-partners, which would result in jets and leptons, with the LSP eluding detection and manifesting as missing transverse energy [26]. To date, no super-partners have been detected at the LHC, suggesting that if SUSY exists, it might manifest at energy scales beyond the LHC's probing capacity [1, 27]. Despite these challenges, the exhaustive exploration of SUSY's parameter space, which includes 120 parameters, is essential for a comprehensive evaluation of the theory [28].

\vspace{-1em}
\section{4. NEUTRINO MASSES}

Neutrinos are among the most abundant particles in the universe, outnumbering electrons and protons by approximately a billion to one. Proposed by Pauli in 1930 to explain the continuous energy spectrum observed in beta decays, the concept of neutrinos suggested the emission of an additional particle alongside the electron [29]. The electron neutrino's discovery several decades later was followed by observations from the Homestake experiment in the 1960s, which recorded only a third of the anticipated solar neutrino flux, leading to the solar neutrino problem. This discrepancy laid the groundwork for the hypothesis of neutrino oscillation among three distinct flavors (electron, muon, tau) with different masses, a theory confirmed four decades later by the Sudbury Neutrino Observatory (SNO), which detected all neutrino flavors, thereby addressing the flux deficit [30]. The existence of nonzero neutrino masses challenges the Standard Model's assumption of massless neutrinos, raising the question of the origin of neutrino masses. The primary theories include the Sterile Neutrino, the Seesaw Mechanism, and the Majorana Neutrino.

\subsection{4.1 The Sterile Neutrino}
Chirality, a quantum mechanical property of particles, exists in two forms: left-handed and right-handed. In the Standard Model, only particles exhibiting both chiralities gain mass through the Higgs mechanism, which involves chiral symmetry breaking and mass generation upon interaction with the Higgs field. Currently, only left-handed neutrinos have been observed, implying their inability to acquire mass via the Higgs mechanism. Sterile neutrinos, theorized as low-mass right-handed particles, interact with ordinary matter predominantly through gravity, making them exceedingly difficult to detect [32]. Their discovery would not only confirm neutrinos as ambidextrous particles but also provide a natural explanation for their mass acquisition through the Higgs mechanism and position them as potential dark matter candidates.

\subsection{4.2 The Seesaw Mechanism}
The Seesaw Mechanism offers an explanation for the neutrinos' lightness by proposing that each Standard Model neutrino flavor has a corresponding heavy right-handed partner. Representing each neutrino flavor with a 2x2 mass matrix, the mechanism posits that the product of the two eigenvalues (corresponding to the Standard Model and heavy neutrino masses) is constant. Consequently, a large mass eigenvalue for the heavy neutrino results in a minuscule mass eigenvalue for the Standard Model neutrino, hence the mechanism's name [33]. Recent findings, such as those from the MiniBooNE experiment at Fermilab, suggest the existence of a fourth neutrino type, supporting the Seesaw Mechanism's predictions~[34].

\subsection{4.3 The Majorana Neutrino}
While right-handed sterile and heavy neutrinos remain undetected, the Standard Model already includes a right-handed neutrino variant, the antineutrino. Majorana particles, hypothesized to be neutrinos that are their own antiparticles, would acquire mass through interactions with their antineutrino counterparts, necessitating a new Higgs-like mechanism beyond the Standard Model Higgs field [35]. The Neutrinoless double beta decay, where a nucleus emits two electrons without neutrinos due to the annihilation of Majorana neutrinos, serves as a potential detection method. The Germanium Detector Array (GERDA) experiment, aimed at observing this decay, has yet to find conclusive evidence, with its second phase ongoing and expected to conclude in the coming years [37].

\section{5. Conclusion}
The observation of phenomena beyond the Standard Model (BSM) underscores the Standard Model's (SM) inherent limitations, revealing it as an incomplete theory of fundamental physics. This review aimed to encapsulate three pivotal BSM phenomena—dark matter, the Higgs boson's unexpected lightness, and neutrino masses—alongside their theoretical underpinnings and the prospects for their empirical detection.

Dark matter, constituting the majority of the universe's matter, remains an enigmatic presence, with hypotheses ranging from modified Newtonian dynamics (MOND) to dark galaxies. However, Weakly Interacting Massive Particles (WIMPs) emerge as the leading candidates within the plethora of dark matter models.

The Higgs boson's discovery stands as a landmark achievement of the 21st century, fulfilling a half-century-old prediction. Yet, the SM falls short of accounting for the Higgs boson's mass sensitivity to radiative corrections. Supersymmetry, an extension that doubles the SM's particle spectrum, offers a plausible resolution to this conundrum.

Neutrinos, ubiquitously present throughout the universe, were shown to oscillate and thus possess mass approximately two decades ago, challenging the SM's massless neutrino premise. Theories explaining neutrino mass invariably introduce a requisite right-handed chiral partner for the SM neutrinos.

In summary, while the SM represents the most triumphant framework in describing fundamental particles and their interactions to date, it does not furnish a comprehensive portrayal of the universe. This realization propels the scientific community towards conducting a wide array of experiments aimed at uncovering new physics and evaluating the validity of existing hypotheses. With Run 3 of the Large Hadron Collider (LHC) already underway, the scientific community is actively probing energy scales up to four times greater than those accessed in its inaugural run, marking a significant stride towards new physics discovery and potentially addressing some of the most pressing questions in contemporary particle physics.

\end{document}